\documentclass[10pt, conference]{IEEEtran} %, compsocconf

\usepackage{graphicx}
\usepackage{float}
\usepackage{amssymb}
\usepackage{array}
\usepackage{enumerate}
\usepackage{booktabs}
\usepackage{verbatim}
\usepackage{fancyvrb,relsize}

\usepackage{xcolor}
\definecolor{labelcolor}{cmyk}{0.22,0.10,0.10,0.10}
\definecolor{listbackgroundcolor}{cmyk}{0.10,0.10,0.05,0.05}

\usepackage{tikz}
\usetikzlibrary{arrows,shapes,snakes,automata,backgrounds,petri}
\usetikzlibrary{fit}
\usetikzlibrary{positioning}
\usetikzlibrary{decorations}
\usetikzlibrary{decorations.markings}
\usepackage{times}
\usepackage{amsmath, amssymb}
\usepackage{latexsym}
\usepackage{graphics}
\usepackage{epsfig}
\DeclareMathAlphabet{\mathsl}{OT1}{ptm}{m}{sl}

\input{epsf}

\newcommand{\ifsubmit}[1]{}

\newcommand{\etal}{{et al.}}

\newcommand{\be}{\begin{itemize}}
\newcommand{\ee}{\end{itemize}}
\newcommand{\bn}{\begin{enumerate}}
\newcommand{\en}{\end{enumerate}}
\newcommand{\bc}{\begin{center}}
\newcommand{\ec}{\end{center}}
\newcommand{\bl}{\begin{flushleft}}
\newcommand{\el}{\end{flushleft}}
\newcommand{\bq}{\begin{quote}}
\newcommand{\eq}{\end{quote}}
\newcommand{\beq}{\begin{equation}}
\newcommand{\eeq}{\end{equation}}

\newcommand{\bmp}{\begin{minipage}}
\newcommand{\emp}{\end{minipage}}

\newtheorem{definition}{Definition}
\newcommand{\bdeff}{\begin{definition}\rm} %\noindent
\newcommand{\edeff}{\end{definition}} %\rule[-0.1mm]{1.0mm}{3mm}

\newtheorem{theorem}{Theorem}

\newtheorem{example}{Example}
\newcommand{\bexa}{\begin{example}\rm} %\noindent
\newcommand{\eexa}{\end{example}}
\newcommand{\eeeexa}{\rule[-0.1mm]{1.0mm}{3mm}\end{itemize}\end{example}}
\newcommand{\eeeeexa}{\rule[-0.1mm]{1.0mm}{3mm}\end{itemize}\end{itemize}\end{example}}

\newcommand{\eenexa}{\rule[-0.1mm]{1.0mm}{3mm}\end{enumerate}\end{example}}
\newcommand{\eeee}{\end{itemize}\end{itemize}}

\newtheorem{axiom}{Axiom}
\newcommand{\baxi}{\begin{axiom}\rm\flushleft}
\newcommand{\eaxi}{\end{axiom}}

\usepackage{ifpdf}
\ifpdf
    \DeclareGraphicsExtensions{.pdf,.png,.jpg}
\else
\fi
\graphicspath{{./}{fig/}{figures/}}  % path for graphics files

\newcommand{\speciallabelsize}{\normalsize\rm}
\newenvironment{exfacts}[1]{\begin{list}{{\speciallabelsize \theenumi.}}{\usecounter{enumi}
        \settowidth{\labelwidth}{{\speciallabelsize #199}}
        \setlength{\leftmargin}{\labelwidth}
        \addtolength{\leftmargin}{2.0\labelsep}}}{\end{list}}
\newcounter{facts}
\newcommand{\bfact}{\begin{exfacts}{F}\setcounter{enumi}{\value{facts}}\renewcommand{\theenumi}{F\arabic{enumi}}}
\newcommand{\efact}{\setcounter{facts}{\value{enumi}}\renewcommand{\theenumi}{
\arabic{enumi}.}\end{exfacts}}

\newcounter{claims}
\newcommand{\bclaim}{\begin{exfacts}{C}\setcounter{enumi}{\value{claims}}\renewcommand{\theenumi}{C\arabic{enumi}}}
\newcommand{\eclaim}{\setcounter{claims}{\value{enumi}}\renewcommand{\theenumi}{
\arabic{enumi}.}\end{exfacts}}

\newcommand{\fbf}{\textbf}

\newcommand{\fsf}[1]{\textsf{\small{#1}}}

\newcommand{\fsc}{\textsc}

\newcommand{\msf}{\mathsf}
\DeclareMathAlphabet{\mathsl}{OT1}{ptm}{m}{sl}

\newcounter{ar-claim}
\newcommand{\barcl}{\begin{exfacts}{ARC}\setcounter{enumi}{\value{ar-claim}}\renewcommand{\theenumi}{$\fbf{AR-claim}_\fbf{\arabic{enumi}}$}}
\newcommand{\earcl}{\setcounter{ar-claim}{\value{enumi}}\renewcommand{\theenumi}{
\arabic{enumi}.}\end{exfacts}}

\newcounter{req}
\newcommand{\breq}{\begin{exfacts}{R}\setcounter{enumi}{\value{req}}\renewcommand{\theenumi}{$\mathit{Req}_{\arabic{enumi}}$}}
\newcommand{\ereq}{\setcounter{req}{\value{enumi}}\renewcommand{\theenumi}{
\arabic{enumi}.}\end{exfacts}}

\newcounter{contribution} 
\newcommand{\bcontr}{\begin{exfacts}{MYC}\setcounter{enumi}{\value{contribution}}\renewcommand{\theenumi}{$\fbf{C}_{\arabic{enumi}}$}}
  \newcommand{\econtr}{\setcounter{contribution}{\value{enumi}}\renewcommand{\theenumi}{
      \arabic{enumi}.}\end{exfacts}}

\usepackage{wasysym}

\begin{document}

\title{The Meaning of Requirements and Adaptation}

\author{
 \IEEEauthorblockN{Amit K.~Chopra}
 \IEEEauthorblockA{
% Department of Information Engineering and Computer Science (DISI)\\
 University of Trento, Italy\\
%  Via Sommarive 14, Povo, Italy\\
 \fsf{chopra@disi.unitn.it}}}

%\institute{Amit~K. Chopra \at 
%  Department of Information Engineering and Computer Science \\
%  University of Trento\\
%  Via Sommarive 14 Povo 38123 Italy\\
%  \email{chopra@disi.unitn.it}
%}

\date{\today}

\maketitle

\pagestyle{plain}
\thispagestyle{plain}
\pagenumbering{arabic}

\begin{abstract}
  The traditional understanding of stakeholders requirements is that
  they express desirable relationships among phenomena in the relevant
  environment.  Historically, software engineering research has tended
  to focus more on the problems of modeling requirements and deriving
  specifications given requirements, and much less on the meaning of a
  requirement itself.  I introduce new concepts that elucidate the
  meaning of requirements, namely, the designated set and the
  falsifiability of requirements.

  By relying on these concepts, I (i) show that the adaptive
  requirements approaches, which constitute a lively and growing field
  in RE, are fundamentally flawed, (ii) give a sufficient
  characterization of vague requirements, and (iii) make the
  connection between requirements modeling and the Zave and Jackson
  sense of engineering.  I support my claims with examples and an
  extensive discussion of the related literature.  Finally, I show how
  adaptation can be framed in terms of Zave and Jackson's ontology.

\end{abstract}

\section{Introduction}
\label{sec:intro}

Self-adaptive systems has developed into a major research theme over
the last decade.  In recent years, the idea that requirements
engineering has a major role to play in the development of this theme
has been gaining ground.  In addition to broad reflections on the
topic \cite{cheng:adaptation:2009}, several technical approaches and
frameworks have been proposed.  In this paper, I want to focus on what
I term the \emph{adaptive requirements} approaches.  I circumscribe
this set to contain all and only those approaches which have at their
heart at least one of the following claims.

\begin{LaTeXdescription}
\item[Flexible requirements claim.] Self-adaptive systems, that is,
  systems that are able to work effectively even when their
  operational environments changes \emph{require} a new class of
  requirements, which I refer to here as \emph{adaptive requirements}.
  In contrast to traditional requirements, which are prescriptive in
  nature, adaptive requirements would be ``flexible'' by definition.
  Whereas prescriptive requirements are either satisfied or violated,
  adaptive requirements would offer more nuanced notions of
  satisfaction.  Adaptive requirements may well be traded off against
  each other by the system at runtime.  Literature that takes this
  position quite clearly includes
  \cite{cheng:adaptation:2009,sawyer:adaptation:2010,whittle:relax:2010,baresi:fuzzy-goals:2010,qureshi:adaptation:2011}.
\item[Requirements at runtime claim.] At design time, there would be
  uncertainty about many things, including the operational environment
  and requirements themselves.  To handle uncertainty effectively, the
  system would need to reason about requirements at runtime when more
  information would be available.  In this sense, requirements
  engineering ceases to be solely an offline activity; systems also do
  it at runtime.  Literature that takes this position quite clearly
  includes
  \cite{berry:adaptation:2005,cheng:adaptation:2009,sawyer:adaptation:2010,whittle:relax:2010,baresi:fuzzy-goals:2010,qureshi:adaptation:2011}.
  \end{LaTeXdescription}

  These two claims amount to a radical departure from the traditional
  requirements engineering, where stakeholder requirements are
  understood to be prescriptive and engineering is understood as an
  activity performed by humans in order to produce systems that meet
  stakeholder requirements.  Because of their radical nature, it would
  be worth investigating whether the claims have any merit and this is
  precisely my purpose in this paper.

  I show that both of these claims are ill-founded.  My approach in
  this paper is as follows.  I introduce two novel concepts that
  concern the fundamental nature of requirements: the \emph{designated
    set} and \emph{falsifiability}.  They are both technical
  contributions in their own right.
  
\begin{LaTeXdescription}
\item[Designated Set.] System engineering is done on the basis of a
  \emph{designated} set of requirements, which is a set of
  requirements which if met would make the stakeholder ``happy''.
  Each requirement in the designated set is of equal, prescriptive
  status: distinctions between requirements, such as critical versus
  noncritical and optional versus mandatory, and preferences over
  requirements make no sense in the set.

\item[Falsifiability.] I elaborate on the traditional idea that
  requirements express stakeholder-designated relationships among
  environmental phenomena.  The elaboration specifically concerns the
  idea that one (especially, the stakeholder) should be able to
  determine whether a requirement is satisfied or violated by
  observing the environment in which the system is running.  If one
  can determine its satisfaction, we term it \emph{satisfiable}; if
  one can determine its violation, we say that the requirement is
  \emph{falsifiable}.  Vague requirements are generally considered
  low-quality requirements in the literature.  I formulate a
  \emph{sufficient} condition for determining a requirement vague: if
  it is both nonsatisfiable and nonfalsifiable.
\end{LaTeXdescription}

I then apply the above two concepts toward making the following
further contributions.

\be
\item I use the concepts as criteria for judging the adaptive
  requirements approaches and show that the adaptive requirements
  approaches fail either one or both of the criteria.
\item Far from rejecting the role of RE in modeling adaptive systems,
  I show that it is possible to conceptualize adaptation in Zave and
  Jackson's conceptual framework for requirements.  In contrast to the
  adaptive requirements approaches, my approach does not resort to
  anything intrinsically adaptive in itself.  And because I frame
  adaptation in terms of concepts that are at the heart of RE, I
  consider this contribution as saying something essential about the
  meaning of adaptation from an RE perspective.
\item The concept of the designated set helps make a high-level
  connection between requirements modeling languages and requirements
  engineering in the sense of Zave and
  Jackson~\cite{zave:dark-corners:1997}, a hitherto dark corner in the
  literature.  
\ee

\subsection{Organization} 

The rest of the paper is organized as follows.
Section~\ref{sec:traditional-RE} discusses the bases of traditional RE
that are relevant for this paper.  Section~\ref{sec:designated-set}
discusses the notion of the designated set and the connection between
requirements modeling languages and engineering.
Section~\ref{sec:falsifiability} introduces the notions of
satisfiability and falsifiability and based on then formulates the
notion of a requirement.  Section~\ref{sec:problems} discusses various
adaptive requirements approaches in the literature and shows their
shortcomings.  Section~\ref{sec:adaptation} then formalizes adaptation
remaining completely within the bounds of traditional RE.
Section~\ref{sec:discussion} is a broad discussion of related
literature.  Section~\ref{sec:conclusions} concludes the paper with a
summary of contributions.

\section{Background}
\label{sec:traditional-RE}

Requirements represent the desirable changes, usually improvements,
that stakeholders wish to see in some existing sociotechnical system.
We understand this sociotechnical system as the operational
environment relevant to the requirements.  The desired changes could
be just about anything: improved functionality, automating some
currently manual task, better storage and recall, more accuracy, more
efficiency, and so on.  In simple terms, requirements express what
stakeholders want.  Without loss of generality, I consider two kinds
of actors in an engineering activity: stakeholders from whom the
requirements come, and engineers who do the system-building in
accordance with stakeholder-expressed requirements.  I also assume
that stakeholders would contract with engineers for meeting their
requirements~\cite{kotonya:requirements:1998}.  Below, I discuss the
three strands of work that led naturally to my observations in this
paper.

\subsection{Zave and Jackson's Engineering}

Requirements engineering, in the sense of Zave and Jackson
\cite{zave:dark-corners:1997}, begins with a set of requirements $R$.
They state clearly that every requirement in $R$ is a constraint over
environment phenomena that the stakeholders want met.  For Zave and
Jackson, engineering refers specifically to the process of identifying
the domain assumptions and coming up with a specification that under
the domain assumptions satisfies the requirements.  The specification
is of a \emph{machine} that can suitably control the environment.  In
other words, I abide by Zave and Jackson's famous conceptualization of
requirements engineering, that is, $K,S\vdash R$, where $K$, $S$, and
$R$ are the set of domain assumptions, the specification, and the set
of stakeholder requirements, respectively.

For the purpose of this paper, we consider $R$ as the \emph{problem}
and any $(K,S)$ pair that satisfies $R$ its solution.  Finding a
suitable $(K,S)$ amounts to exploring the solution space.  Zave and
Jackson state that RE for any set of requirements is \emph{completed}
when we have found a suitable $(K,S)$: from then on, it's just largely
a technical matter of implementing the $S$.  Hence, in this paper, we
don't distinguish between specifications and their implementations.
Jackson's problem frames approach \cite{jackson:problem-frames:2000}
embodies essentially this idea of engineering.

Engineering is an inherently creative enterprise.  Once an engineer is
given a $R$, to identify the $K$ and $S$, he or she would draw upon
his expertise and domain knowledge.  He or she would also need to
study the environment in which the system will be deployed.  If $R$
turns out to be infeasible upon further analysis, for example, if
there were no cost-effective solution for $R$, then the engineer and
the stakeholder need to go back to the drawing board to identify a new
set of requirements.  

Requirements engineering as a field means many things, including
elicitation, modeling, management, coming with specifications, and so
on.  For the purposes of this paper, however, when I use the term
`requirements engineering', I mean it specifically in the Zave and
Jackson sense: that one (attempts to) come up with a specification
given some requirements.

\subsection{Goal Modeling}

Before an engineer can go to work in the above sense of Zave and
Jackson, he and the stakeholder must identify properly the
requirements.  This step is important because initial stakeholder
expressions may be incomplete, inaccurate, inconsistent, and so on.
Many different kinds of modeling and analysis go in this (see
\cite{vanLamsweerde:requirements:2009} Chapters 2 and 3).  One
analysis technique is via the \emph{elicitation} and \emph{modeling
  and analysis} of stakeholder goals.  In requirements engineering,
this step is often referred to as exploration of the \emph{problem
  space}.

Over the years, goal modeling and analysis has turned out to be an
influential technique for exploring the problem space.  In technical
terms, the literature does not convey a strictly uniform relation of
goals and requirements.  According to van Lamsweerde, a requirement is
a goal for which a single active component---``agent'' in his
terms---is responsible \cite{vanLamsweerde:requirements:2009}.
Accordingly to Mylopoulos {\etal} \cite{mylopoulos:goals:1999}
requirements are represented as goals.  Ant\'{o}n
\cite{anton:goals:1996} expresses an intuition similar to Mylopoulos
{\etal}.  Alternatively, goals are understood as the rationale for
requirements.  Given that the categories of goals are mostly similar
to categories of requirements (for example, the distinction between
hard and soft goals resembles that between functional and
nonfunctional requirements), for the purposes of this paper, I will
assume what should be a fairly uncontroversial reading of the
literature---that goal models express stakeholder requirements using
the constructs of goal modeling (such as AND-OR decomposition,
contributions links, and so on).

Goal modeling has proved to be especially influential in the modeling
and analysis of nonfunctional requirements
\cite{mylopoulos:nonfunctional-requirements:1992} and reasoning about
alternatives for the satisfaction of requirements, broadly,
\emph{variants}
\cite{mylopoulos:goals:1999,giorgini:tropos:2002,vanLamsweerde:requirements:2009,jureta:techne:2010}.

\subsection{Requirements Monitoring}

As pointed out by Fickas and Feather \cite{fickas:monitoring:1995},
even if RE is completed for some requirements and we have deployed an
implementation of the specification, it does not mean that the
requirements play no further role at system runtime.  There is still
immense value in monitoring that requirements are being met by the
system, especially in dynamic environments, where the domain
assumptions may change over time.  More prosaically, it may simply be
that the implementation is erroneous.  In such cases, requirements may
end up being violated, and if they are, then some corrective measure
would need to be taken.  Newer work \cite{souza:awareness:2011}
considers finer-grained aspects of monitoring in the context of
adaptive systems.

Since requirements are what stakeholders care about, it follows that
the ability to monitor, in principle, the \emph{satisfaction} and the
\emph{violation} of requirements is crucial.  As we shall see in
Section~\ref{sec:falsifiability}, the ability to determine
satisfaction is independent of the ability to tell violation.

\section{Designated Requirements}
\label{sec:designated-set}

\bdeff
\label{def:happy-set}
A \emph{happy} set of requirements is a set of stakeholder
requirements such that meeting this set of requirements would imply
that the minimum requirements of the stakeholder have been met.
\edeff

``Minimum'' is to be interpreted like this: imagine there were a
contract between the stakeholder and the engineer.  Then if the
engineer were to build a system that met a happy set of requirements,
then the stakeholder could not hold the engineer liable for violating
the contract.

Let me illustrate this with the help of examples.

\bexa
\label{exa:happy-or} The stakeholder communicates to the engineer that
one of $r_0$ and $r_1$ must be met.  This means that there are three
possible happy sets: $\{r_0\}$, $\{r_1\}$, and $\{r_0,r_1\}$.  The
empty set would not be a happy set for the stakeholder.  \eexa

\bexa
\label{exa:happy-optional} The stakeholder communicates that $r_0$
must be met but $r_1$ is optional.  Here, the happy sets are only two:
$\{r_0\}$ and $\{r_0,r_1\}$; $\{r_1\}$ is not a happy set.
\eexa

Definition~\ref{def:designated-set} introduces the idea of designated
set of requirements.

\bdeff
\label{def:designated-set} A \emph{designated} set of requirements is
a happy set of requirements that the engineer chooses to do
engineering for, that is, come up with a specification for.  \edeff

In other words, the designated set would be Zave and Jackson's $R$
that the engineer decided to come up with a specification $S$ for.  By
the very nature of $R$, all requirements in $R$ are prescriptive (Zave
and Jackson use the term `optative').  Further, they are all of equal
status: the engineering must account for all of them.

Referring to Example~\ref{exa:happy-or}, the engineer could pick
$\{r_0\}$ for engineering.  In doing so, he would elevate $\{r_0\}$ to
the status of the designated set.  Alternatively, he could elevate
either $\{r_1\}$ or $\{r_0,r_1\}$ to that status.

Referring to Example~\ref{exa:happy-optional}, the engineer could
elevate $\{r_0,r_1\}$ to the status of the designated set;
alternatively, he could have elevated $\{r_0\}$ to that status.
Clearly, the stakeholder would prefer that the designated set be
$\{r_0,r_1\}$, because it also includes the optional requirement.  The
choice, however, belongs to the engineer, and if he or she were to
elevate $\{r_0\}$ to the status of the designated set, then that too
would be a perfectly legitimate choice.  By annotating a requirement
optional, the stakeholder has in fact made the choice possible for the
engineer.

The concepts of happy and designated set are important to making the
connection between requirements elicitation and modeling on the one
hand and requirements engineering in the sense of Zave and Jackson on
the other.  The set of happy sets would depend upon the modalities and
relationships (for example, \emph{optional}, \emph{AND-OR
  decomposition}, and so on) the stakeholders use to express their
requirements, which would also be reflected formally in the associated
modeling language.  The engineer would then apply Zave and Jackson's
methodology to one of the happy sets, that is, the designated set.
Next, I illustrate this by taking goal modeling as the language of
expression.

\subsection{Happy and Designated Sets in Goal Modeling}

\begin{figure}[htb!]
\centerline{\includegraphics[scale=.40]{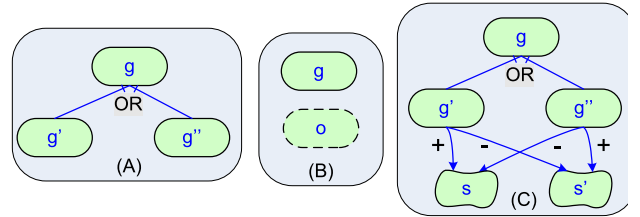}}
\caption{Goals models and alternatives}
\label{fig:goal-models}
\end{figure}

Let's say an engineer draws up the goal model of
Figure~\ref{fig:goal-models}(A) based on stakeholder expressions of
requirements.  In Figure~\ref{fig:goal-models}(A), following commonly
followed goal-modeling semantics, there are three happy sets:
$\{g'\}$, $\{g''\}$, and $\{g',g''\}$.  The engineer can elevate any
one of them to the status of a designated set.

In Figure~\ref{fig:goal-models}(B), $o$ is an optional goal \`{a} la
Techne \cite{jureta:techne:2010}; therefore, there are two happy sets
to choose from: $\{g\}$ and $\{g,o\}$, either of which the engineer
can treat as the designated.  What was \emph{optional} in the goal
model (B) is indistinguishable as such in the designated set: in the
designated set, $o$ is as prescriptive a requirement as $g$ is.

Soft goals complicate the picture somewhat.  Soft goals (for example,
low cost and efficiency) help a stakeholder choose from alternative
requirements, that is,
\emph{variants}~\cite{mylopoulos:nonfunctional-requirements:1992}.
Figure~\ref{fig:goal-models}(C) offers the three variants ($\{g'\}$,
$\{g''\}$, and $\{g',g''\}$) except that \emph{contributions} to two
soft goals $s$ and $s'$ offer additional guidance to the parties
involved (there is no common understanding in the literature of what a
variant formally is; I follow~\cite{chung:nonfunctional:2009}).  In
this particular example, because a positive contribution to $s$ means
a negative contribution to $s'$ and vice versa, it is not clear what
the happy set should be.  Here, the model serves as an aid to the
parties in understanding the relation among requirements and the
various alternatives.  Based on this understanding, further
elicitation and refinement of the goal model may be carried out so
that the happy sets can be identified.  For example, the stakeholder
may express a preference for $\{g''\}$ over the other two.  Hence
there would be only one happy set, that is, $\{g''\}$.  This example
shows that the notion of a happy set, although it overlaps with that
of a variant, is also a distinct one: whereas
Figure~\ref{fig:goal-models}(C) has variants, it has no happy sets.

The above account of goal modeling makes clear the connection between
goal models and requirements engineering: in general, a goal model may
allow for multiple happy sets, which may or may not coincide with the
variants in the goal model.  The engineer can then elevate one of thus
happy sets to the status of a designated set for engineering.  The
connection places goal modeling in a role logically prior to the
engineering: goal modeling help understand the problem space;
engineering ties a selected problem with the solution space.  There is
a lot of work that supports my account, starting from the early work
on nonfunctional requirements analysis
\cite{mylopoulos:nonfunctional-requirements:1992} and including newer
work such as \cite{bryl:req-anal:2009,jureta:techne:2010}.  More
interestingly perhaps, Yu {\etal} \cite{yu:goals-variability:2008}
show how a goal model with many alternatives may be interpreted as a
product line (not any specific product).

(Note that one can give a somewhat simpler account of goal modeling.
Once the variants are identified, the stakeholder must indicate one of
variants to the engineer for engineering.  This account though is
subsumed by my account above: in this account, there would effectively
be only one happy set, that which the stakeholder points out.)

Feather {\etal}~\cite{feather:runtime:1998} use the notion of
alternatives in goal models as the basis for runtime adaptation:
OR-decomposed goals are interpreted as runtime alternatives for
satisfying the parent goal.  I will address that work in more detail
later but let's return to Figure~\ref{fig:goal-models}(A) and consider
a hypothetical scenario.  Let's say the designated set were $\{g'\}$.
Now if at runtime, $g'$ were violated, could the system not
\emph{adapt} by attempting to satisfy the $g''$, and thereby attempt
to satisfy $g$?  After all, both $g'$ and $g''$ are means to the same
end $g$.  The answer is no.  The reason is that the system was
engineered with respect to the designated set $\{g'\}$.

\section{The Falsifiability of Requirements}
\label{sec:falsifiability}

A requirement is an optative statement about the environment.  As
stated above, engineering refers to the process by which we specify a
machine that can suitably control the environment such that the
requirement is met.  We shall refer to such a machine as the
\emph{system that meets the requirement}.  Notice, however, that the
notion of a requirement itself makes \emph{no reference} whatsoever to
any particular specification or its implementation: it is simply what
the stakeholder wants to \emph{observe} in the environment under
consideration.  Either he observes it, in which case we say his
requirement was satisfied, or not, in which case we say it was
violated.

This binary nature of requirement is fundamental to a stakeholder's
\emph{perception} of the quality of a system that meets those
requirements.  If the system violates a requirement, the stakeholder
will perceive the system as low quality (at least with respect to the
requirement); if it meets the requirement, as high quality.  However,
a requirement that gives no basis for him to make such a judgment is
itself low-quality.  Requirements that are described as \emph{vague}
are an example of low-quality requirements.  In RE, the criticism is
often from the engineer's perspective: it is not clear how to
implement vague requirements.  However, the real problem with vague
requirements is borne out from the stakeholder's perspective: how to
tell the satisfaction or violation of a vague requirement.  The reason
is that given a vague requirement an engineer can always choose an
\emph{arbitrary} interpretation of the requirement (meaning without
stakeholder involvement) and build a system that meets it.  For
example, he may interpret the ``quickly'' in ``goods will be ordered
quickly'' as ``two days''.  However, that means the engineer is making
business decisions that are properly the stakeholder's prerogative,
which is clearly undesirable.  Gordijn and Akkermans
\cite{gordijn:value-based-re:2003} make a similar observation in the
context of value propositions.

\subsection{The Satisfiability and Falsifiability of Requirements}

Consider requirements for a purchase order handling system.  Consider
the requirement~\ref{req:goods-twodays}.

\breq \item \label{req:goods-twodays} Goods shall be ordered within
two days of an authorized request being placed for them.  \ereq

If a requested goods are ordered within two days, we may say that
\ref{req:goods-twodays} is satisfied.  However, we have to qualify
that claim a little: we can claim satisfaction only ``for that
instance'' of goods ordering.  We cannot claim the general
satisfaction of the requirement because the number of ordering
instances is unbounded.  In the rest of this paper, I will use the
term \emph{satisfiable} to mean that we can at least tell the
satisfaction of particular instances at runtime.

However, if ordering takes more than two days for any instance, then
we say that \ref{req:goods-twodays} is violated in general.  We say
that a requirement is \emph{falsifiable} if at runtime we can tell if
the requirement was violated; else, it is nonfalsifiable.  We say that
\ref{req:goods-twodays} is both \emph{satisfiable} and \emph{falsifiable}.

Consider the alternative requirement~\ref{req:goods-percent}.  This
requirement is similar, in the aspects pertinent to the discussion
here, to one found the IEEE Standard dealing with software
requirements specifications \cite{ieee:srs-standard:1998}.

\breq \item \label{req:goods-percent} Goods shall be ordered within
two days of the order being placed in 60\% of the instances.  \ereq

``60\% of the instances'' by itself is a meaningless constraint
because the space of instances for calculating the percentage is not
specified.  Therefore, given \ref{req:goods-percent}, one can conclude
\emph{neither} its satisfiability nor falsifiability.  If
\ref{req:goods-percent} were instead ``60\% of the instances each
month'' (\ref{req:goods-percent-modified}) then it would be both
satisfiable and falsifiable.

\breq
\item \label{req:goods-percent-modified} Goods shall be ordered within
  two days of the order being placed in 60\% of the instances each
  month.  \ereq

  Requirements \ref{req:goods-asap}--\ref{req:fairness} below are
  neither satisfiable nor falsifiable.  Following conventional RE
  terminology, they would be deemed \emph{vague}.  However, the notion
  of vagueness itself has not been formulated precisely in the
  literature.  Being neither satisfiable nor falsifiable implies
  vagueness.

  \[\lnot \msf{falsifiable}(r) \land \lnot
  \msf{satisfiable}(r)\Rightarrow \msf{vague}(r)\]

\breq 
\item \label{req:goods-asap} Goods shall be ordered \emph{as soon as
    possible} after they have been requested.
\item \label{req:throughput} The system shall have \emph{high}
  throughput.
\item \label{req:fairness} All requests shall be handled
  \emph{fairly}.  \ereq

  \ref{req:goods-eventually} is satisfiable but not falsifiable.  Like
  \ref{req:goods-asap}, it places no time limit on the ordering of
  goods; however, the moment goods are ordered, it is satisfied.

\breq 
\item \label{req:goods-eventually} Goods requested shall be
  \emph{eventually} ordered.  
  \ereq

  The problem with requirements that are not falsifiable (satisfiable
  or not) is not that they are not stated formally.  Later, we shall
  consider approaches where \emph{as soon as possible} is formalized
  in a fuzzy logic.  And clearly, the notion of \emph{eventually} has
  a formal meaning in temporal logic.  The problem is simply that they
  yield no criterion by which to judge their violation.

\subsection{Requirements, Formally}

Logically, a requirement expresses a constraint over predicates that
describe the environment.  We imagine an observer (could be the
stakeholder) who evaluates the constraint by observing the
environment.  Below, $t,t'$ etc. are variables over real time since we
are dealing with observations.

\ref{req:goods-twodays} can be expressed by the following constraint.

\[\forall(x,t,t')\;\mathit{requested}(x,t) \land \lnot
\mathit{ordered}(x,t') \rightarrow \mathit{diff}(t',t)<2D\]

This means that if goods instance $x$ has been requested at time $t$
and at a time $t'$ such that the difference $t$ and $t'$ is greater
than 2 days, one observes that the book has still not been ordered,
the requirement is violated.  Hence this requirement is falsifiable.
However, for some value of $t'$ within two days, if
$\mathit{ordered}(x,t')$ is observed to be true, then the requirement
is satisfied for that instance.  Hence \ref{req:goods-twodays} is also
satisfiable.

The \ref{req:goods-eventually} may be expressed as the following.

\[\forall(x,t)\exists t'\;request(x,t) \rightarrow ordered(x,t')\]

If for some value of $t'$, $ordered(x,t')$ is observed to be true;
then \ref{req:goods-eventually} is satisfied for this instance; hence,
this requirement is satisfiable.  However, this requirement is not
falsifiable because for any value of $t'$ where $\lnot ordered(x,t')$
is true, there is a later time $t''$ such that $ordered(x,t'')$ may
turn true.  Thus, at no time, can this requirement be considered
violated.

This brings us to the crux of the matter: formally, I interpret a
requirement as a constraint that separates the legal states of the
environment (where it is not violated) from the illegal ones (where it
is violated).  If the set of illegal states is empty, then the
requirement is not falsifiable.  If however, the set of legal states
is empty, then it is not satisfiable.  An inconsistent requirement is
not satisfiable.  For example, ``the temperature in the room shall
always be $\ge 18^{\circ}\mathrm{C}$ and $< 18^{\circ}\mathrm{C}$'' is
not satisfiable: all environment states are illegal.

We would all agree that from the point of view of requirements
quality, nonsatisfiable requirements are clearly bad; however,
nonfalsifiable requirements are worse.  The reason is that an engineer
would be unable to engineer a system to meet a nonsatisfiable
requirement; therefore at some point, he will likely return to the
stakeholder for a revised requirement.  However, as discussed earlier,
if the stakeholder specifies a nonfalsifiable requirement, an engineer
may interpret it arbitrarily.  In the worst case, he may not even
implement it, and yet, at runtime, stakeholders would never be able to
claim violation of the requirement.  Moreover, an engineer could not
easily be held accountable for the system he engineered on the basis
of stakeholder-given nonfalsifiable requirements.

\ref{req:goods-asap}--\ref{req:fairness}, without additional
elaboration, have no reasonable interpretation as constraints that
separate the legal from the illegal.  \ref{req:goods-twodays} is a
falsifiable and satisfiable elaboration of \ref{req:goods-asap}.
\ref{req:throughput-fs} and \ref{req:fairness-fs} are falsifiable and
satisfiable expressions of \ref{req:throughput} and
\ref{req:fairness}, respectively.  (In particular,
\ref{req:fairness-fs} is violated when a request that came after some
other request is served before the latter.)

\breq
\item\label{req:throughput-fs} The system shall service at least 200
  outstanding requests per day.
\item\label{req:fairness-fs} Requests will be served on a FIFO
  basis. \ereq

\subsection{Comments on Some Classes of Requirements}

Some requirements express that something must \emph{eventually}
happen, as in \ref{req:goods-eventually}.  We list below some more
examples.  Notice that sometimes ``eventually'' does not explicitly
appear in the statement of the requirement, for example, in
\ref{req:train-eventually}.

\breq
\item\label{req:train-eventually} Upon applying the emergency brakes,
  the train must come to a halt.
\item \label{req:light-eventually} The traffic light shall eventually
  turn green.
\ereq

For each of \ref{req:train-eventually} and \ref{req:light-eventually},
the time period during which the requirement is to be met would be
important to the stakeholder.  It is important to railway designers
that the train halt within some specified number of seconds for safety
purposes (perhaps depending on the speed bracket).  It matters to
traffic planners whether the traffic light turns green within 30
seconds or 60.  \ref{req:goods-eventually} shows that such
requirements are not limited to safety-critical systems.

Some requirements, e.g., \ref{req:escalator-instantaneous}, are meant
to be satisfied instantaneously.  

\breq
\item\label{req:escalator-instantaneous} The escalator must start when a
  person steps onto it and stop when all persons have stepped off it.
\ereq

I interpret instantaneously as \emph{for the same value of time}, in
other words, for the same observation.
\ref{req:escalator-instantaneous} is violated if in some observation,
there is a person on the escalator, but the escalator hasn't started.

Traditionally, functional requirements are understood to be about
system behavior whereas nonfunctional requirements are understood to
be about the quality of the system.  Traditional examples of
nonfunctional requirements include efficiency, usability, security,
and so on.  In my discussion of satisfiability and falsifiability, I
did not have to distinguish between them.  Whether functional or
quality, each requirement must be satisfiable and falsifiable.  For
quality requirements, this means that they should be adequately
metricized.  \ref{req:throughput-fs} may be considered a metricized
version of \ref{req:throughput}.  More interestingly perhaps,
\ref{req:goods-twodays} may be said to combine both functional and
quality concerns.

Fickas and Feather \cite{fickas:monitoring:1995} introduce
\ref{req:vague-license} in their well-known work on requirements
monitoring.

\breq
\item\label{req:vague-license} Users should not have to wait
  \emph{unduly} long to get a license.  
\ereq

As it is, this requirement is neither satisfiable nor falsifiable; it
is vague.  To Fickas and Feather's credit, at another place in the
paper, they mention \emph{within parentheses} that
\ref{req:vague-license} is monitored by watching users who have
applied for licenses for some \emph{predetermined amount of time}.
Such broken-up descriptions of requirement obscure their nature.
Choosing five days as the value for predetermined amount of time, the
requirement would have been better stated as \ref{req:fixed-license},
which is a falsifiable requirement.

\breq
\item\label{req:fixed-license} Users should not have to wait more than
  five days to get a license.  \ereq

\section{The Problems with Adaptive RE Approaches}
\label{sec:problems}

I discuss some of the adaptive RE approaches. These approaches either
fail at least one of the criteria below.

\begin{LaTeXdescription}
\item[Criterion of the Designated Set] Any approach that talks about
  system engineering from requirements should clearly identify the
  designated set.
\item[Criterion of Complete Observability] Every requirement in the
  designated set should be both satisfiable and falsifiable.
\end{LaTeXdescription}

Notice that the criterion of complete observability does not say every
requirement should be falsifiable; it limits itself to requirements in
the designated set.  This implies, for example, that in goal models
themselves, soft goals (which are used to represent nonfunctional
requirements) need not necessarily be metricized.  However, if the
exercise of goal modeling leads to a designated set (as explained the
in Section~\ref{sec:designated-set}), then every requirement in it,
whether functional or nonfunctional, should be both satisfiable and
falsifiable.

\subsection{Flexible Requirements}

Whittle {\etal}~\cite{whittle:relax:2010} introduce a language
\fsc{relax} with new operators for the express purpose of supporting
flexible requirements.  The presumed benefit of flexible requirements
is that it gives the system room to act flexibly.  For example, if in
a situation, only one from among a prescriptive and a flexible
requirement can be satisfied, then the system can forgo the
satisfaction of the latter.  More generally, an adaptive requirement
can be satisfied to a lesser or greater degree depending upon the
circumstances.

Let's consider some examples Whittle {\etal} give to illustrate how
\fsc{relax} would work in a smart home setting.

\breq
\item\label{req:alarm} The system shall raise an alarm if no activity by Mary is detected for some hours (to be decided) during normal waking hours.
\item\label{req:fridge} The fridge shall detect and communicate with all food packages.
\item\label{req:fridge-relaxed} The fridge shall detect and
  communicate with \fsc{as many} food packages \fsc{as possible}.
  \ereq

  According to Whittle {\etal}, the requirement \ref{req:fridge} is
  prescriptive.  Requirement \ref{req:fridge-relaxed} is its relaxed
  version: instead of communicating with \emph{all} food packages,
  only \fsc{as many as possible} need to be communicated with.  If the
  requirements specification were
  \{\ref{req:alarm},\ref{req:fridge-relaxed}\}, then in situations
  where all available resources were required to satisfy
  \ref{req:alarm}, the system would forgo the satisfaction of
  \ref{req:fridge-relaxed} or satisfy it to a lesser degree.  Besides
  quantitative relaxed requirements expressed using operators such as
  \fsc{as many as possible}, one can also express temporal relaxed
  requirements using operators such as \fsc{as early as possible},
  \fsc{as late as possible}, and so on.

  Whittle {\etal} want to avoid expressing environmental conditions as
  part of requirements in order that the system would have the freedom
  to adapt as best suits the satisfaction of the requirements.
  Whittle {\etal} allow that downstream designers and programmers
  could implement relaxed requirements as they see best.

In \fsc{relax}, one can express both prescriptive and relaxed
requirements.  Whittle {\etal} make an important semantic distinction
between them.  The former are understood as critical, the latter as
noncritical.  Whittle {\etal} provide a methodology for identifying
requirements that can potentially be relaxed.

There are two problems with \fsc{relax}.  One, by separating the
requirements into critical and noncritical, it violates the criterion
that all requirements are of equal prescriptive status.  More
specifically, consider that the downstream designers could interpret a
relaxed requirements as they see fit.  For example, \fsc{as soon as
  possible} could be interpreted by them as two minutes or twenty
minutes or four hours---they are not constrained in their choice.
However this is exactly the kind of arbitrary interpretation of
requirements that I criticized earlier.  And in this, \fsc{relax}
fails the criterion of the designated set; it fails because it is not
clear that the stakeholders would be happy with the designers' choice.
(In Section~\ref{sec:adaptation}, I discuss how the critical versus
noncritical distinction can be accounted for without violating the
criterion.)

Two, the \emph{noncritical}, that is, relaxed requirements fail the
criterion of complete observability: they are neither satisfiable nor
falsifiable.  They are in fact vague.  Consider their explanation of
\fsc{as early as possible} $\phi$: ``$\phi$ becomes true in some state
as close to the current time as possible'' but ``technically allows
$\phi$ to become true at any point after the current time''.  Any
point after current time means any time in the future.  Clearly, this
requirement is not falsifiable.  Is the requirement satisfiable?  One
could argue that it is satisfiable at all times.  But then the
requirement \fsc{as early as possible} $\phi$ would be exactly
\fsc{eventually} $\phi$, which is satisfiable but not falsifiable, and
hence not vague.  The reason I deem it vague lies in the formalization
of the relaxed requirements as fuzzy sets: one can only claim a degree
of satisfaction with the requirement.  You may consider that a
stakeholder would be satisfied at degree 100, but would he be
satisfied at degree 99\ldots60\ldots30\ldots1\ldots?  The answer is
not clear, and hence the requirement is vague.

  Further, even the critical requirement ``eventually $\phi$'' fails
  the criterion of complete observability.  Specifically, it fails
  falsifiability (as I have already discussed in
  Section~\ref{sec:falsifiability}).

  Baresi {\etal} \cite{baresi:fuzzy-goals:2010} introduce a language
  with operators similar to \fsc{relax} to support runtime adaptation.
  Qureshi {\etal} \cite{qureshi:adaptation:2011} advocate flexible
  requirements similar to \fsc{relax}.

\subsection{RE at Runtime}

Recall that the designated set is a happy set of stakeholder
requirements which a system has been to designed to meet.  Now, a
system that has been designed to meet a set of requirements cannot
possibly reason about them.  It was the engineer who reasoned about
the requirements, looking at it as a problem-solving exercise, in
coming up with the specification, which the system is implemented to
conform to.  This reasoning necessarily happens offline.  Therefore,
to claim that a system can reason about its requirements at runtime
would be tantamount to claiming that a solution can reason about the
problem it solves---clearly, an absurd notion.

It is necessary to address the issue of uncertainty that seems to be
one of the key motivations for adaptive requirements approaches.
Adaptive requirements approaches presume that new specification
languages that explicitly support uncertainty are
required~\cite{cheng:adaptation:2009}.  Two kinds of statements about
uncertainty appear in the literature: one, there is uncertainty about
the requirements in the sense that it is not possible to
\emph{anticipate} them all a priori, and two, there is uncertainty
about how a system's operational environment will be.  The adaptive
requirements approaches claim these uncertainties could be tackled by
reasoning about requirements at runtime.

Engineering a system to meet a set of designated requirements
necessarily means building the system to meet a set of
\emph{anticipated} requirements.  As I argued above, a system cannot
reason about its own requirements, let alone unanticipated ones.  

It is true that a system's operational environment can change in
unpredictable ways.  However, there is no way to resolve this problem
except by studying and analyzing the possible contingencies and
building a system to support the most reasonable (perhaps, after a
cost-benefit analysis of some sort) contingencies.  A system cannot
adapt to a contingency it was not built to handle.  There is no
substitute for sound engineering.  Section~\ref{sec:adaptation}
emphasizes this and shows the logical construction of an adaptive
system from an RE point of view.

Below, I discuss specific works that subscribe to the RE at runtime
view, all of which fail the criterion of the designated set for
various reasons.

\subsection{Reasoning about Goals}

Feather {\etal}~\cite{feather:runtime:1998} claim two complementary
approaches for dealing with violations of requirements: one, by simply
designing the system better to take care of contingencies, and two, by
the system itself making ``acceptable changes to the requirements'' at
runtime.  However, it follows from the criterion of the designated set
that it is not possible for the system to make any changes to
requirements: it is simply built to the designated set.  Nonetheless,
it is interesting to consider what makes Feather {\etal} make the
claim.  Feather {\etal}'s approach is goal-oriented.  The system has
some goals that may potentially be satisfied by multiple alternative
refinements.  What Feather {\etal} mean by ``acceptable change'' is
that some alternative goal refinement may be adopted to ensure
satisfaction of the system's top-level goals.  As is common in goal
modeling, Feather {\etal} identify the goals with requirements.  The
adoption of alternate refinements at runtime leads Feather {\etal} to
claim that the system is making acceptable changes to requirements.
The mistake though is in identifying system goals with stakeholder
requirements.  The system goals are correctly seen not as requirements
but as a declarative program to a planning-based execution engine.
The actual requirements of this system would have been something else
altogether: they are what lead to the system's representation in terms
of goals. That system representations are not the same as requirements
actually follows from Zave and Jackson's work: stakeholder
requirements may be far removed (but connected to by domain
assumptions) from the specifications to which programs are written to
conform to.  In hindsight, I see my own work on goals and agent
adaptation as having the sense of Feather {\etal}
\cite{chopra:goals-com-aamas:2010}.

% Goals-based reasoning and planning has roots in artificial
% intelligence and some researchers in the field of requirements
% engineering seek to apply the techniques developed there to goals
% modeling as is prevalent in RE.  From the statements that ``goals
% represent requirements'' (from RE) and ``goals can be reasoned about
% and planned for at runtime'' (from AI), one is tempted to conclude
% ``therefore, requirements can be reasoned about at runtime''.  This is
% false conclusion.  Because in AI, 

\subsection{User Input versus Stakeholder Requirement}

Qureshi {\etal} \cite{qureshi:req-runtime:2011} state, ``the system
playing the role of the ``analyst'', performs requirements elicitation
and analysis (i.e. refinement and update of adaptive requirements
specification) itself in order to provide solutions to satisfy
end-user's needs in a particular context by looking up and selecting
the appropriate services. Thus, the requirements refinement is dynamic
in that it is realized as a service selection process''.  However,
this reflects a profound confusion between what is \emph{input} for a
system versus its requirements.  Staying with the service selection
analogy, a user-given constraint on the types of services a system
needs to select is \emph{input} to the system.  The \emph{stakeholder
  requirement} presumably was that either the system shall find
services according to user-input constraints or announce failure if it
cannot find them in some specified amount of time.  Inverardi and Mori
\cite{inverardi:req-runtime:2011} succumb to the same confusion.

To say stakeholder requirements for a system are the same as user
inputs to the system is to make a grave category error.

\subsection{Diluting the Nature of Requirements Engineering}

Compared to Zave and Jackson's characterization of requirements
engineering as a creative problem-solving enterprise, Berry {\etal}
\cite{berry:adaptation:2005} present a weaker characterization of RE.
According to them, doing RE means determining ``the kinds of inputs a
system may be presented'' and ``the system's responses to these
inputs''.  Berry {\etal} use this characterization of RE to justify
their claim that a dynamically adaptive system (DAS) does \emph{RE at
  runtime} because it switches behavior depending on the input.  They
miss the crucial part about ``determining'', which is where creativity
is involved.  It seems that they mistook Parnas and Madey's
characterization of software requirements as an input-output relation
\cite{parnas:system-model:1995} for the activity that produces the
relation.  Further, by their characterization, there would be few
systems in the world that are not DAS: wouldn't a simple program with
an if-else statement conditioned on some environmental expression be a
DAS?  Berry {\etal} \cite{berry:adaptation:2005} disclaim RE at
runtime in the sense of strong AI, which renders their choice of the
terms ``RE at runtime'' doubly puzzling.

\subsection{Optional and Preferred Requirements}

There is a sense that many in the RE community have that the system
has some flexibility in meeting requirements, at least in the sense
that some requirements are \emph{optional} or \emph{preferred} and the
system can dynamically choose among them.  Although, I claimed above
that in the designated set, every requirement is equal and
prescriptive, my claim requires deeper introspection in light of some
recent work in this area.

Jureta {\etal} \cite{jureta:ontology:2008} make no claims related to
software adaptation.  However, they extensively discuss optional and
preferred requirements and offer an excellent starting point for
discussing them further.  Jureta {\etal} think it a shortcoming of
Zave and Jackson's $K,S\vdash R$ formulation that it does not account
for \emph{optional} requirements and go on to reformulate the $R$ in
$K,S\vdash R$ to include optional requirements as a first-class
concept, couched in what they call \emph{attitudes}.  However, as I
pointed out earlier, whereas the concept of optional requirements does
have a role to play in goal modeling and analysis, once the designated
set is identified from among the various alternatives, it is simply a
set with each requirement of equal, compulsory status.  Consider that
if one found a $(K,S)$, one would want to know whether it's a solution
of just the compulsory requirements or if it also includes a subset of
the optional ones.  However, ``wanting to know'' means solving the
engineering problem where those optional requirements are considered
compulsory.  In other words, the engineering problem is always solved
for a set of compulsory requirements, which is what the designated set
captures.  The IEEE Standard dealing with software requirements
specifications \cite{ieee:srs-standard:1998} also supports the idea of
optional requirements explaining them as something that enables
engineers to go beyond what is required of them.  However, consider
that if a magnanimous engineer does decide to satisfy some of the
stakeholder's optional requirements, that implies he or she has
elevated their status to compulsory as far as finding a solution is
concerned.  It is also just as reasonable that if an engineer had to
come up with a solution for a problem containing both compulsory and
optional requirements, he would save himself trouble by simply not
considering the optional requirements.

A set of preference relations over requirements is the other part of
Jureta {\etal}'s attitudes.  I agree completely with Jureta {\etal}
that preference relations help stakeholders choose among requirements.
However, again, just as I argued for optional requirements above, Zave
and Jackson's engineering is done for a chosen, that is, designated,
set of requirements.  The relations \emph{optional} and
\emph{preferred} make no sense over requirements in the designated
set.  So, in essence, at least from the point of view of optional and
preferred requirements, Zave and Jackson's formulation is perfectly
fine.

Since there are no optional or preferred requirements at runtime, it
makes no sense to talk about adaptation with respect to them.  A flaw
of Qureshi {\etal}'s ``core'' ontology for self-adaptive systems
\cite{qureshi:adaptation:2011} is that it is based on Jureta
{\etal}'s.  Further, requirements that we deem ``optional'' and
``preferred'' at runtime can be explained the same way I explain the
critical-noncritical distinction in Section~\ref{sec:adaptation}.

\section{Addressing Adaptation: Two Ideas of Switching}
\label{sec:adaptation}

In the foregoing, I have shown that the adaptive requirements
approaches are conceptually flawed.  In this section, I show that it
is possible to reason about adaptation by applying Zave and Jackson's
work.

\subsection{Switching Machines}

Let $R$ be a set of requirements.  Let

\[\mathit{Sol}=\{(K_i,S_i)|0\le i \le n\:\: \msf{and}\: K_i,S_i\vdash R
\}\]

That is, $\mathit{Sol}$ is a set of solutions (traditionally, the set
of solutions is a singleton).  Methodologically, the set of solutions
is obtained by varying the domain assumptions, selecting those that
engineering will need to account for.  This step involves significant
analysis, for example, taking into account the probability of the
operating environment mirroring particular domain assumptions, doing a
cost-benefit analysis of engineering the system to work under selected
assumptions, and so on.  Given some $R$, identifying a
\emph{reasonable} set of solutions for it by considering variations of
the domain assumptions is part of \emph{good engineering}.

Imagine a controller for a set of solutions $\mathit{Sol}$ that
monitors the environment to figure out which $K$ current environmental
conditions mirror and puts into operation the corresponding machine
$S$.  We say that $(K,S)$ is the \emph{state} of the controller.  We
say that the controller \emph{performs an adaptation} when it switches
from $(K,S)$ to some other $(K',S')\in\mathit{Sol}$.  (If the
controller determines environmental conditions that mirrors no $K$ in
$\mathit{Sol}$, then it continues to operate in the current state.  We
imagine a $\mathit{null}$ machine for the initial state.)

I consider the controller to be a generic artifact. So logically, I
consider an adaptive system that meets $R$ to be characterized by
$\mathit{Sol}_R$, that is, the set of $R$'s solutions.  Let us keep in
mind that here $R$ is fixed, and we are simply \emph{switching
  machines}.  Definition~\ref{def:machine-switch} gives a formal
characterization of a machine-switching system.

\bdeff
\label{def:machine-switch}
A machine-switching system is characterized by the tuple $\langle
\mathbf{K},\mathbf{S}, \mu \rangle$, where 

\be
\item $\mathbf{K}$ is a set of sets of domain assumptions,
\item $\mathbf{S}$ is a set of machines, and
\item $\mu:\mathbf{K}\rightarrow \mathbf{S}$ is a bijection
\ee
\edeff

We can characterize a machine-switching system by a set of pairs, each
of the form $(K,S)$.

\subsection{Switching Requirements}

Ali {\etal} \cite{ali:context:2010} consider requirements themselves
as \emph{contextual}; in other words, certain requirements apply only
in certain situations.  Can we account for that in the traditional
framework?  The answer is yes.

In contrast to the ideas of switching machines, let us consider the
case where the requirements themselves change depending on changes in
the operational environment.  To distinguish from domain assumptions,
we term these environmental conditions \emph{modes}.  Thus one can say
that if the mode $E_0$ holds, then the system should behave according
to the requirements $R_0$, if $E_1$, then $R_1$, and so on.  In other
words, whereas the requirements were fixed in the case of switching
machines, here the requirements themselves change depending on the
operational environment.

That the system behave according to particular requirements depending
on the current mode is itself a requirement, which we term a
\emph{modal requirement}.  The modal requirement is in fact the
complete stakeholder requirement.  Let $\mathcal{C}$ be a modal
requirement.  Then $\mathcal{C}$ is essentially a case statement of
the form $E_0::R_0, E_1::R_1,\ldots, E_n::R_n$.  Now for each $R_i$
($0\le i\le n$), following the notion of solution set described above,
let $M_i$ be its machine-switching system.  Now imagine a
mode-controller that monitors the environment and switches to the
fulfillment of a particular set of requirements depending on the
current mode.  This means that it switches to the corresponding
machine-switching system.  That is, when $E_i$ holds, the
mode-controller will put into operation $M_i$.  The machine switching
system's controller further matches the environmental conditions with
the domain assumptions to select a particular machine to put into
operation.  That is, if $\mathit{Sol}_i=\{( K_0,S_0),( K_1,S_1),
\ldots, ( K_m,S_m)\}$, then if $K_j$ ($0\le j\le m$) holds, then $S_j$
is put into operation.

Since the mode-controller is generic, the system is characterized as
in Definition~\ref{def:req-switch}.

\bdeff 
\label{def:req-switch} A mode-switching system is a tuple $\langle
\mathbf{E},\mathbf{M}, \xi\rangle$ where \be
\item $\mathbf{E}$ is a set of sets of environment conditions,
\item $\mathbf{M}$ is a set of machine-switching systems,
\item $\xi: \mathbf{E}\rightarrow \mathbf{M}$ is a bijection
\ee
\edeff

We can characterize a mode-switching system by a set of pairs, each of
the form $(E,M)$.  Definition~\ref{def:equivalence} states what it
means for two adaptive systems to be equivalent.
Theorem~\ref{thrm:equivalence} then states the equivalence between
machine-switching systems and mode-switching systems.

\bdeff
\label{def:equivalence} 
Two systems are equivalent for a given set of environment variables
if and only if they both put the same machine into operation for every
environmental condition defined over the variables.  
\edeff

\begin{theorem}
\label{thrm:equivalence}
Let $\mathbf{M}=\{\{(K_0^0,S_0^0), (K_1^0,S_1^0),\ldots, (K_j^0,S_j^0)\},$ \\
$\{(K_0^1,S_0^1), (K_1^1,S_1^1),\ldots,
(K_k^1,S_k^1)\},\ldots,\{(K_0^i,S_0^i), (K_1^i,S_1^i),$ \\
$\ldots,(K_l^i,S_l^i)\}\}$ be a set of $i$ machines, labeled
$M_0,M_1,\ldots,M_i$.  The mode-switching system
$\{(E_0,M_0),(E_1,M_1),\ldots, (E_i,M_i)\}$ is equivalent to the
machine-switching system $\{(K_0^0\cup E_0,S_0^0), (K_1^0\cup
E_0,S_1^0),\ldots,(K_j^0\cup E_0,S_j^0), (K_0^1\cup E_1,S_0^1),
(K_1^1\cup E_1,S_1^1),\ldots, (K_k^1\cup E_1,S_k^1),\ldots,(K_0^i\cup
E_i,S_0^i), (K_1^i\cup E^i,S_1^i),\ldots, (K_l^i\cup E^i,S_l^i)\}$.
\end{theorem} {\bf Proof.} It follows from the explanation above of
how mode-switching works.

The proof of the theorem is simple but the theorem is significant.  It
shows that the mode-switching system reasons no more about
requirements than a machine-switching system, which itself does not
reason about requirements since they appear nowhere in its
characterization.  Further, I did not have to extend Zave and
Jackson's formulation to formalize adaptation; I simply had to apply
it.

The potential benefit of considering multiple machine systems versus
single machine systems is modularity.  This is an area that requires
further investigation.

\subsection{Accounting for the Critical-Noncritical Distinction}

The idea behind mode-switching, which as you have seen above, is one
of expression, not of meaning.  However, it can be used to account for
the effect researchers seem to be grasping after when they talk of
requirements being considered critical versus noncritical and optional
versus mandatory at runtime.

Consider a mode-switching system $\langle \mathbf{E},
\mathbf{R},\xi\rangle$.  What Whittle {\etal} deem a critical
requirement is better understood as a requirement that is applicable
in all $E\in\mathbf{E}$; in other words, it appears in all
$R\in\mathbf{R}$.  A noncritical requirement is a requirement that is
applicable in some $E$s (presumably those in which the system is
functioning largely as expected).  For example, in the smart home
example, I could express the requirement that in the course of
\emph{normal} operation (a mode), modeled as $E$, both \ref{req:alarm}
and \ref{req:fridge} apply.  That may be considered the default case.
In the case of \emph{abnormal} operation though (another mode),
modeled as $E'$, \ref{req:fridge} does not apply.

\section{Discussion}
\label{sec:discussion}

I discuss some additional literature and then conclude the paper with
a summary of my claims and arguments.

\textbf{Understanding Requirements.} Zave and Jackson
\cite{zave:dark-corners:1997} and Letier and van Lamsweerde
\cite{letier:requirements:2002} rule out specifications that contain
unbounded future references such as ``eventually''.  However, they
both allow such statements as stakeholder requirements.  The criterion
of falsifiability for stakeholder requirements rules out such
statements even as stakeholder requirements.

Parnas and Madey's REQ relation \cite{parnas:system-model:1995}
between monitored and controlled environment variables is similar in
nature to a specification since it refers to a ``controller''.  In
other words, Parnas and Madey are describing the $S$ in $K,S\vdash R$,
not the stakeholder requirements $R$.  By contrast, in this paper,
when discussing satisfiability and falsifiability, I refer to $R$.
Zave and Jackson's contribution is the $K,S\vdash R$ relation and the
idea (and a corresponding ontology) that in requirements engineering,
we are principally concerned with phenomena in the environment.  One
of my contributions in this paper concerns a more precise
understanding of the nature of $R$.

Tackling the challenge of what the terms in the expression of a
requirement mean in the real world is an important
one~\cite{zave:dark-corners:1997}.  However even those requirements in
which the meaning of the terms in clearly established may turn to be
nonfalsifiable or worse vague.

\textbf{Requirements Quality.}  That requirements themselves be of
high-quality is often stressed.  van Lamsweerde
\cite{vanLamsweerde:requirements:2009} list the following qualities:
(1) completeness, (2) consistency, (3) adequacy, (4) pertinence, (5)
good structuring, (6) traceability, (7) modifiability, (8)
feasibility, (9) comprehensibility, (10) unambiguity, and (11)
measurability.  Qualities (1)--(8) are described as relations between
multiple objects, so they are not intrinsic to the notion of a
requirement. (9) refers to intelligibility, so its purpose is to
enforce a good syntax. (10) refers to designations (discussed above).
Measurability lumps together not only testability and verifiability
(more below), but also that users be able to monitor whether
requirements are being satisfied or not during system operation.  Some
of the \emph{fit criteria} van Lamsweerde introduces later in the book
as a way of making requirements measurable are vague: they involve
calculation of percentages over unbounded spaces.

The notion of fit criteria is due to Robertson and
Robertson~\cite{robertson:fit-criteria:2006}.  They seem to be
motivated by similar concerns as are addressed by falsifiability: one
should be able to tell whether a requirement is met or not.  They also
state that the fit criterion of a requirement \emph{is} the
requirement, that is, the fit criterion says something essential about
the meaning of a requirement.  Further, they agree that the
stakeholder must agree about the fit criterion for any requirement.
However, just as in van Lamsweerde's book, some of examples of fit
criteria Robertson and Robertson give involve calculation of
percentages over unbounded spaces.  Further, they seem to
unnecessarily limit the significance of their work when they say that
``the tester ensures that each of the product's requirements complies
with the fit criterion''.  This would imply that whether requirements
are met could be determined by testing before deployment.  This is not
true in general because as I explained in
Section~\ref{sec:falsifiability}, for most requirements, we would be
able to claim satisfaction only for specific instances (which is what
testing would do), not for the requirement itself.

Some may insist that a requirement is \emph{whatever} a stakeholder
wants.  In that case, even if the stakeholder wants and insists on a
nonsatisfiable or nonfalsifiable requirement, the fact remains that it
will be such a low-quality requirement as to have nothing of the
nature of a requirement.

\textbf{Formal Verification.}  Given a formal model of the domain
assumptions and the specification, one can, in principle, statically
verify if the model satisfies a formally-expressed requirement (for a
use of formal verification, see \cite{atlee:requirements:1996}).
Notice that in standard temporal logic, using the operations
$\mathbf{G}$ (\emph{always}) and $\mathbf{F}$ (\emph{eventually}) one
can express the following liveness property.

\[\mathbf{G}(request(x)\rightarrow \mathbf{F}\:ordered(x))\] 

One can also verify them against suitable state-machine-based models.
The above temporal logic property is quite similar to
\ref{req:goods-eventually}, and verification algorithms can tell
whether or not the Kripke structure satisfies the property.  Real-time
model checking formalisms may be used to verify properties that encode
that a particular event happen within a particular time limit of
another (similar to \ref{req:goods-twodays}).  Such verification would
increase our \emph{confidence} that the system will meet requirements
at runtime, in other words, that the system actually serves its
intended purpose.  However, no matter how extensive the verification,
one cannot guarantee that at runtime the requirements will be met.
The value of testing is similar: it also increases confidence in the
system.

The IEEE standard considers a requirement verifiable if there
exists a finite cost-effective procedure for determining whether a
system implementation meets the requirement.  According to the
standard, \ref{req:goods-percent}, a vague requirement, would be
\emph{verifiable}.  Stated as it is, it is unclear on what basis the
standard claims it verifiable.

\textbf{Switching Behavior.} The idea of switching machines is similar
to Zhang and Cheng's idea of switching programs
\cite{zhang:adaptive-software:2006}; programs in their terminology
would be machines in ours, and the transition would occur upon changes
in the domain properties.  Salifu {\etal} \cite{salifu:switching:2007}
conceptualize adaptation as switching machines when the domain
assumptions change, in much the same way I do.  However, Salifu
{\etal} consider detecting the changes in the environment and ensuring
that the appropriate machine is being used as additional RE problems.
It would be interesting to further investigate this difference.

Requirements can be specified flexibly: one can specify the boundaries
within which the variables concerning the requirements should remain.
If the values stray outside the boundaries, then the requirement is
violated.  Such requirements being falsifiable would be fine. For
example, one can say that the temperature in the room shall always be
greater than $18^\circ$ but less than $24^\circ$.  This leaves room
for an intelligent controller to behave adaptively, as in Epifani
{\etal}~\cite{epifani:adaptation:2009}.

\section{Conclusions}
\label{sec:conclusions}

Let me summarize my contributions.

\bn
\item System engineering is done on the basis of a designated set
  of requirements, which is the a set of requirements which if met
  would satisfy the stakeholder.  The concept of the designated set is
  crucial in making the connection between requirements modeling
  languages and engineering in the sense of Zave and Jackson.
  Further, each requirement in the set is of equal, prescriptive
  status.  In this set, there is no such thing as critical versus
  noncritical requirements, or optional versus mandatory requirements,
  or preferences over requirements.

\item I showed that requirements ought to be falsifiable, otherwise,
  in principle, stakeholders would never be able to tell when a
  requirement were violated, and in practice, they would be
  dissatisfied.  Further, I characterized requirements that are both
  nonfalsifiable and nonsatisfiable as vague requirements.  I also
  showed that falsifiability stands distinct from the notions
  discussed in the literature.

\item I showed how a simple application of Zave and Jackson's seminal
  work can be used to model and reason about adaptation.  In doing
  this, I also accounted for the idea of contextual requirements.

\item What ties together the above three contributions is my
  evaluation of adaptive requirements approaches.  Some of the
  adaptive requirements approaches fail the criterion of a designated
  set whereas others fails complete observability; some both.  The
  third contribution shows that it is possible to remain within the
  bounds of traditional RE and yet come up with meaningful
  adaptation-related ideas.  \en

  To support my claims, I have provided many examples and conducted an
  extensive survey of the literature.  If one accepts my
  characterization of requirements as constraints that are both
  falsifiable and satisfiable by observations of the environment, then
  he or she must concede the flaws in the notion of flexible
  requirements.  And if one concedes that a system is built to a meet
  a designated set of requirements, then there is no sense in talking
  about a system reasoning about its requirements, let alone doing any
  RE at runtime.  One may be temped to claim that the confusions in
  the literature that I alluded to earlier are merely about
  terminology.  I think that confusions about terminology often stem
  from a deeper misunderstanding about the nature of things, in this
  case, about requirements.

  My evaluation of the adaptive requirements approaches is principled.
  It is not based upon ``local'' observations about particular
  adaptive requirements approaches; instead, I point our their
  shortcomings based on the two criteria that have only to do with the
  nature of requirements, not with adaptation.

  Many in the RE community believe that requirements have an important
  role to play in the engineering of adaptive systems.  That belief is
  not unreasonable.  Work on monitoring requirements and their
  relation to feedback loops~\cite{souza:awareness:2011}, parameter
  adjustment within bounds~\cite{epifani:adaptation:2009}, and so on
  are interesting directions.  Section~\ref{sec:adaptation} is my own
  contribution in support of this belief.  My evaluation in this paper
  is limited to the \emph{adaptive requirements} approaches, that is,
  those that claim at least one of the following: (1) requirements
  themselves are flexible, and (2) a system can reason about
  requirements at runtime.

  The adaptive requirements approaches, especially those that advocate
  flexible requirements, start with the \emph{unjustified} premise
  that self-adaptive systems need a new kind of requirements and
  engineering.  If a self-adaptive system is one that switches
  behavior depending upon the environmental conditions, then I showed
  how one can build such systems even remaining strictly within the
  confines of traditional RE.  If a self-adaptive system is one that
  engineers, evolves, changes, or reasons about requirements at
  runtime, I showed that this is conceptually impossible, for
  solutions cannot reason about the problems they are solutions of.
  ``RE at runtime'' is a meaningless term.  Unfortunately, it is also
  misleading.

  From relatively modest intellectual beginnings in
  \cite{feather:runtime:1998,berry:adaptation:2005}, which have over
  time proved influential, the adaptive requirements theme has gained
  in visibility and credibility.  This is evidenced by papers in
  respected peer-reviewed venues
  \cite{sawyer:adaptation:2010,whittle:relax:2010,baresi:fuzzy-goals:2010,qureshi:adaptation:2011}
  and a widely-cited Dagstuhl seminar report on self-adaptive systems
  that practically enshrines flexible requirements as the way
  forward~\cite{cheng:adaptation:2009}.  There have been two workshops
  on the theme of requirements at runtime at the \emph{Requirements
    Engineering} conferences of 2010 and 2011.  'Uncertainty' is the
  theme for RE 2012 and one of the tracks there is titled `RE at
  Runtime'.  The new \emph{Software Engineering for Self-Adaptive
    Systems} (SEAMS) symposia prominently feature adaptive
  requirements-related work.  Given the positive momentum the adaptive
  requirements theme has and the relatively bold claims it makes, it
  is worth examining it with a critical eye.

  I would like to emphasize one final thing.  It seems to me that much
  of requirements modeling and analysis is presented as relatively
  agnostic to the parties involved in the activity, namely, the
  stakeholders and engineers, with much more emphasis placed on
  technical aspects---for example, new modeling constructs, the notion
  of a variant, and associated algorithms and their performance.
  Whereas the technical themes are interesting, taking a more explicit
  party-oriented approach could lead to new insights in the field.  I
  did some of that in this paper.  I gave an example where going from
  goal models to the designated set needed communication among the
  stakeholders and engineers.  In fact, a happy set is ``happy'' from
  the stakeholder's perspective, whereas a designated set combines
  both the engineer's and the stakeholder's perspectives.
  Falsifiability is motivated from the stakeholder's perspective.  I
  also dwelled briefly on how both the notions of falsifiability and
  designated set relate to stakeholder satisfaction with the execution
  of work (contract) by the engineer.  In general though, we need to
  do much more, for instance, model explicitly the communication
  between the parties that lead to the requirements and the related
  contracts being set up.  A communicated-based normative account of
  requirements seems to me an exciting way forward.

% The most alarming thing of all about the adaptive requirements
% approaches is that the stakeholder is completely missing from the
% discourse.  One cay say that what I did in this paper was argue from
% the contractual perspective: the designated set of requirements is the
% set that the stakeholder wants the engineer to meet and it is the
% stakeholder who cares about falsifiability because that tells him how
% good a job the engineer has done.  Adaptive requirements approaches
% instead take a purely technical perspective.  

%\section*{Acknowledgments}  This paper owes a great deal to engaging
%discussions with Fabiano Dalpiaz, Vitor Souza, Alexei Lapouchnian,
%Raian Ali, Olga Gadyatskaya, Adina Sirbu, Neil Ernst, Feng-Lin Li,
%John Mylopoulos, Paolo Giorgini, Munindar Singh, and Michael Jackson.
%The research was supported by a Marie Curie Trentino Cofund
%fellowship.

%\bibliographystyle{IEEEtrans}
%\bibliography{Amit}

% Generated by IEEEtranS.bst, version: 1.13 (2008/09/30)

\end{document}